\documentclass[fleqn,10pt]{wlscirep}
\usepackage[utf8]{inputenc}
\usepackage[T1]{fontenc}
\usepackage{lineno}

\title{Ultra-compact topological photonic crystal rainbow nanolasers operating in the 1550 nm telecom band with wavelength-scale mode volumes}

\author[1,$\dag$]{Feng Tian}
\author[1,$\dag$]{Yilan Wang}
\author[1,$\dag$]{Wendi Huang}
\author[2,*]{Xuan Fang}
\author[1]{Shengqun Guo}
\author[1,*]{Taojie Zhou}
\affil[1]{School of Microelectronics, South China University of Technology, Guangzhou, 510641, China}
\affil[2]{State Key Laboratory of High Power Semiconductor Lasers, School of Physics, Changchun University of Science and Technology, Changchun, 130022, China}

\affil[*]{corresponding authors: Xuan Fang (fangx@cust.edu.cn), Taojie Zhou (taojiezhou@scut.edu.cn)}

\affil[$\dag$]{these authors contributed equally to this work}

\begin{abstract}
Density-integrated, multi-wavelength nanoscale lasers with ultra-low power consumption and ultra-compact footprints are essential for energy-efficient, fast and high-throughput data processing. Currently, on-chip multi-wavelength lasers predominantly rely on arrays of discrete large-scale conventional semiconductor lasers that are susceptible to the fabrication imperfections. Topological rainbow nanolasers, which spatially confine and emit specific topologically protected light frequencies, offer a prospective approach for achieving ultra-compact integrated multi-wavelength light sources with enhanced robustness against perturbations and defects. However, it remains a significant challenge to achieve highly localized topological rainbow trapping in nanocavities for laser emission with both high quality factors and ultra-small mode volumes. Here, we experimentally report ultra-compact topological photonic crystal rainbow nanolasers operating in the 1550 nm telecom band. Specifically, we present rainbow-like emission with uniform wavelength spacing and wavelength-scale mode volume $\sim$ 0.7$(\lambda/n)^{3}$ in a one-dimensional topological rainbow nanolaser, exhibiting robust lasing operation across a wide temperature range and a spectral tuning capability of approximately 70 nm. Additionally, we demonstrate an ultra-compact two-dimensional topological rainbow nanolaser in an exceptionally compact footprint of nearly 0.002 mm$^{2}$, featuring a broad rainbow spectra with 64 continuously tuned lasing peaks. Our work provides a promising method for realizing robust and nanoscale multi-wavelength tunable laser sources, paving the way for numerous potential applications in ultra-compact photonic chips.

\end{abstract}
\begin{document}

\flushbottom
\maketitle

\thispagestyle{empty}


\section*{Introduction}

Nanolasers with ultra-compact footprint, low energy consumption, and high modulation speed have attracted considerable attention as advanced on-chip coherent light sources for next-generation high-density integrated photonic circuits\cite{chen2014nanophotonic,ma2019applications,dong2014silicon,matsuo2018low}. Intensive research have been conducted to develop nanoscale lasers with unprecedented performance based on various cavity types, including photonic crystal (PhC) nanolasers\cite{zhou2020continuous,yang2017room}, nanoridge lasers\cite{shi2017optical}, nanowire lasers\cite{huang2001room} and metallo-dielectric nanolasers\cite{azzam2020ten,ouyang2024singular}. However, a plethora of studies so far have focused on individual device for single wavelength emission. To enable fast and high-throughput data processing, multi-wavelength lasers with distinct frequencies are attractive and crucial for dense integrated photonic systems\cite{deka2020nanolaser,lee2020photonic}, particularly for on-chip wavelength-division-multiplexing and intra-chip optical interconnects\cite{liang2022energy,matsuo2018low}. Currently, the aforementioned applications are in general implemented by integrating arrays of discrete conventional large-scale lasers, such as distributed feedback lasers, which are limited by their large footprint, high power consumption, and sensitivity to fabrication imperfections. Thus, achieving robust and ultra-compact multi-wavelength lasing emission with controllable wavelength spacing is highly desirable to tremendously increase integration density and reduce energy consumption.

Meanwhile, topological photonics has emerged as a rapidly growing field of research, offering a unique route to engineer the flow of light\cite{lu2014topological,ozawa2019topological,khanikaev2017two}. Notably, significant advancements have been made in  developing topological lasers with new functionalities with remarkable robustness against imperfections \cite{zeng2020electrically,bandres2018topological,ota2020active}. Rainbow trapping, which spatially separates and traps different light frequencies\cite{tsakmakidis2007trapped,tsakmakidis2017ultraslow,dixon2020tunable}, has opened new avenues for achieving high-density integrated multi-wavelength nanophotonic devices\cite{elshahat2021perspective}. Rainbow nanolaser, capable of producing a rainbow-like emission in nanoscale, provides a distinctive scheme for high-density light sources in nanophotonic circuits.\cite{dannenberg2021multilayer}. Recently, several theoretical approaches have been proposed combining rainbow trapping effect and topological concepts to design novel devices with topologically protected rainbow confinement of light\cite{lu2021topological,zhao2022topological,zhang2024observation}. Nevertheless, the topological rainbow nanophotonic devices have yet to be fully explored \cite{lu2022chip}, with most current methods constrained by  weak light confinement and low quality factor (Q-factor), hindering their application in high-performance nanophotonic devices particularly in ultra-compact topological rainbow nanolaser.

In this work, we experimentally demonstrate room-temperature low-threshold topological PhC rainbow nanolasers in the 1550 nm telecom band. 
Rainbow-like spectra with controllable wavelength spacing is achieved in a one-dimensional (1D) topological rainbow nanolaser, showing a small mode volume ($V_{m}$) $\sim$ 0.7$(\lambda/n)^{3}$ and a low threshold of $\sim$ 0.9 kW cm$^{-2}$. In addition, the 1D topological rainbow nanolaser exhibits robust lasing operation across a wide temperature scale, presenting a broad spectral tuning range of approximately 70 nm. By integrating spatially separated topological 1D edge modes and 0D corner modes with carefully engineered frequencies, we further demonstrate an ultra-compact 2D topological PhC rainbow nanolaser with 64 continuously tuned lasing peaks in an exceptionally small footprint of approximately 0.002 mm$^{2}$. This work provides a crucial method for realizing ultra-compact topological coherent rainbow light sources, and will have a wealth of potential applications in high-density photonic chips.

\begin{figure}[t!]
	\centering
	\includegraphics[width=0.93\linewidth]{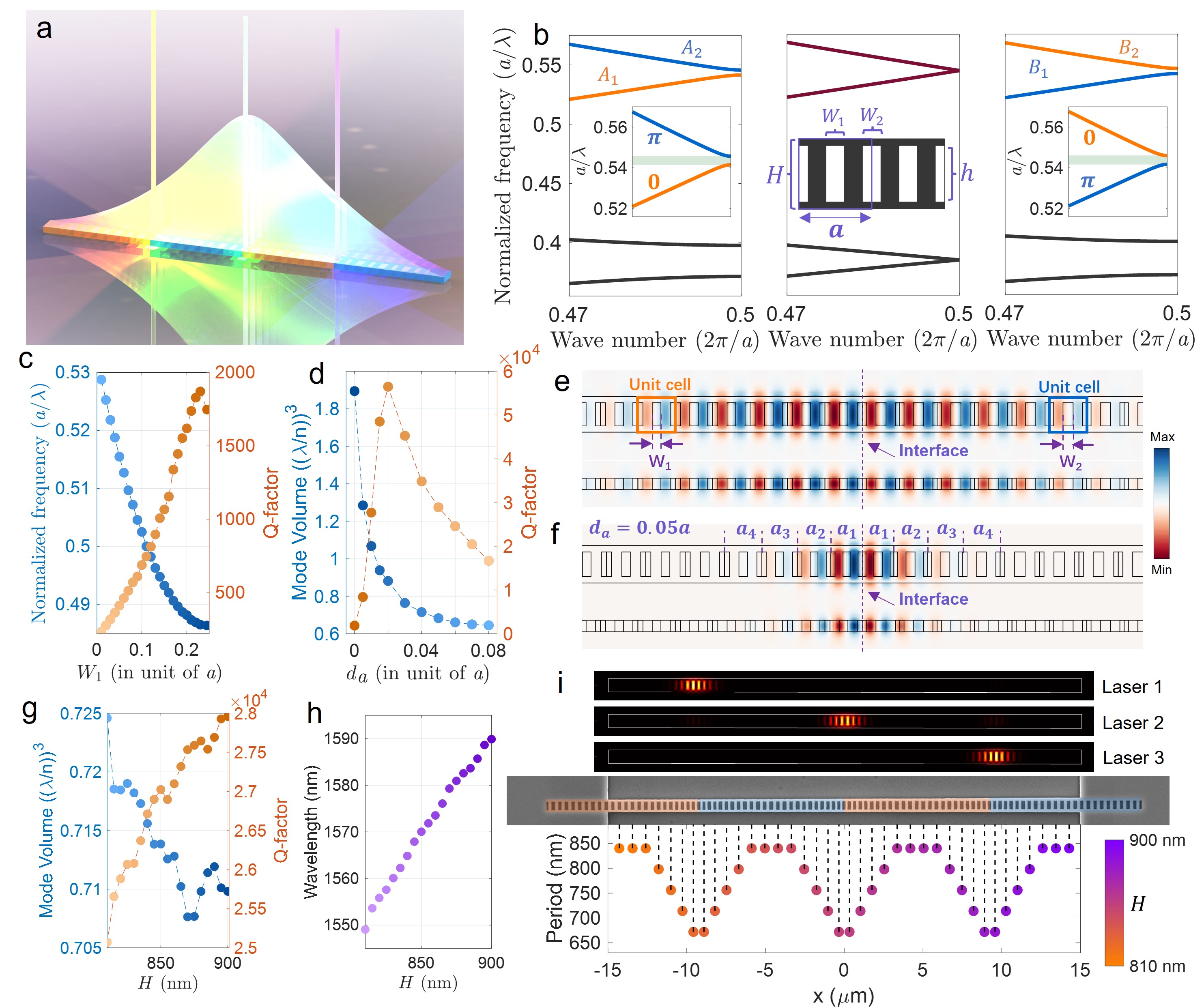}
	\caption{\textbf{1D Topological rainbow nanocavity design concept}. (a) Schematic illustration of the 1D topological rainbow nanolaser. The topologically nontrivial PhC and its trivial counterpart are highlighted in orange and blue, respectively. (b) The left, middle, and right panels present the calculated band diagrams for the unit cell with $W_{1} =$ 0.15$a$, $0.25a$ ($W_{1}$ airhole centered) and $W_{2} =$ $0.35a$ ($W_{2}$ airhole centered), respectively. The inset in the middle panel shows PhC unit cell marked with structural parameters. The insets in the left and right panels show the zoomed diagram of the third and fourth bands, respectively. (c) Simulated normalized frequencies and Q-factors of the topological edge-state mode without tapered structure under various $W_{1}$. (d) Simulated $V_{m}$ and Q-factors of the topological edge-state mode under various $d_{a}$. (e) and (f) show the top and cross-sectional views of the calculated electric field ($E_{y}$) profiles for the 1D topological PhC nanocavity with $d_{a}$ values of 0 and 0.05$a$, respectively. (g) and (h) illustrate the simulated $V_{m}$, Q-factors, and resonant wavelengths for the tapered topological PhC nanocavity with various height ($H$) of nanobeam. (i) The top-view SEM image of the fabricated ultra-compact 1D topological rainbow nanolaser. The corresponding structural parameters are shown below the SEM image. The calculated $E^{2}$ field distributions for three resonant wavelengths (1550 nm, 1570 nm, and 1590 nm) are shown above the SEM image.}
	\label{fig1}
\end{figure}

\section*{Results}

\begin{figure}[t!]
	\centering
	\includegraphics[width=0.93\linewidth]{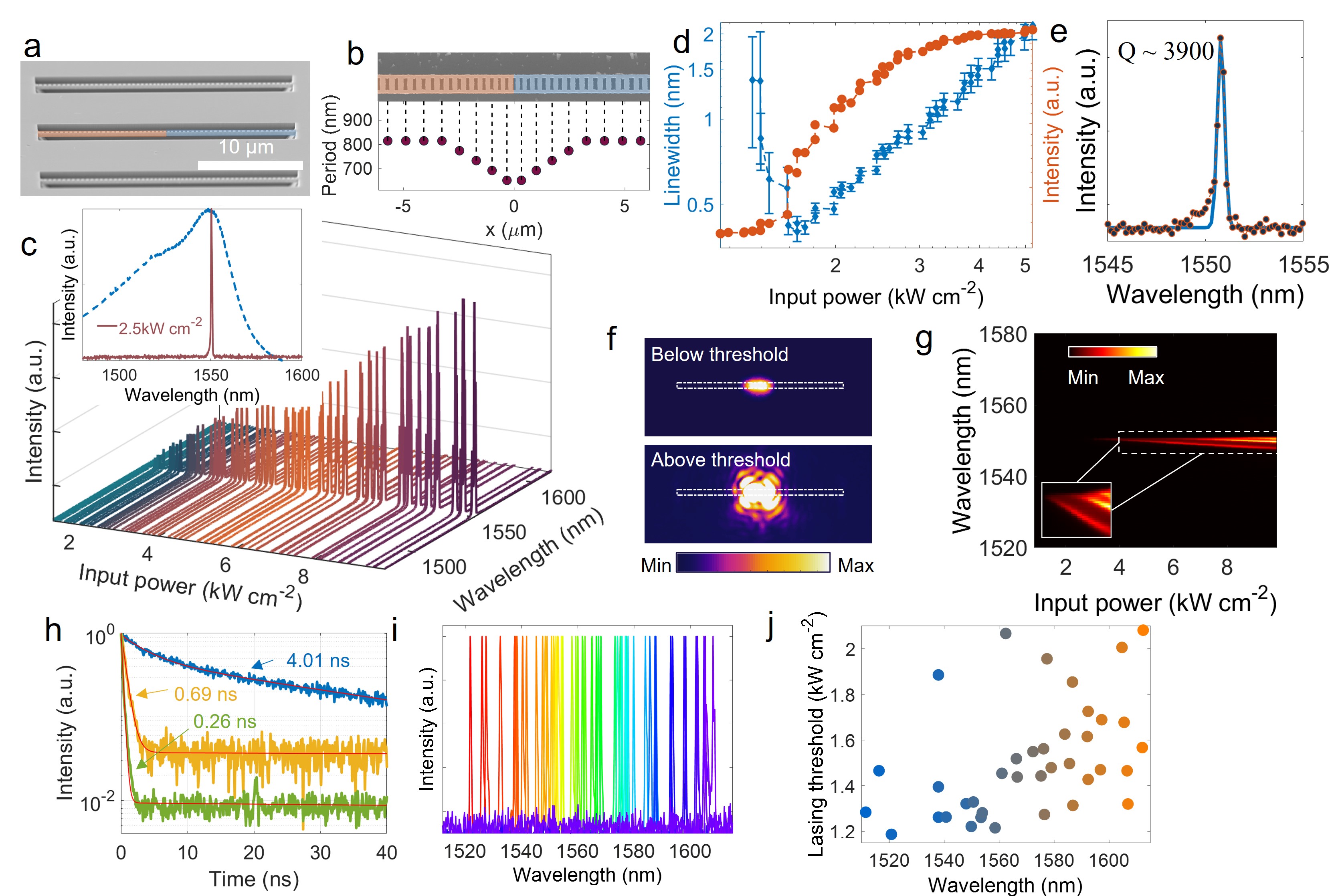}
	\caption{\textbf{Lasing properties of the tapered topological PhC nanolaser}. (a) The tilted SEM image of fabricated nanolaser array. (b) The SEM image of the central tapered nanocavity region, with corresponding periods shown below the image. (c) Collected spectra under various input powers. The inset shows the laser spectra (solid curve) collected at 2.5 kW cm$^{-2}$ and a typical spontaneous emission spectra (dashed curve). (d) Logarithmic plot of the collected $\textit{L}-\textit{L}$ curve and linewidth for the lasing peak around 1550 nm. (e) Gaussian fitting (blue line) of the spectra measured just below the threshold, indicating a linewidth of approximately 0.4 nm. (f) Near-field images of the lasing mode measured below and above the threshold. (g) Image of measured lasing spectra at various input powers. (h)  Normalized TRPL spectra of the unpatterned region (blue curve) and the topological nanolaser pumped slightly below (yellow curve) and above (green curve) threshold. The red curves are fits to a bi-exponential decay model. (i) Normalized lasing spectra and (j) corresponding thresholds of representative single-mode tapered topological nanolasers with various structural parameters.}
	\label{fig2}
\end{figure}

Figure 1(a) presents a schematic of an ultra-compact 1D topological PhC rainbow nanolaser. The structure consists of two suspended PhCs with distinct topologies (depicted in orange and blue), alternatively spliced to form multiple nanocavities at the interfaces. The designed topological rainbow nanobeam is constructed from InGaAsP-based PhC slab with a total thickness of 260 nm and height of $H$. The unit cell contains two rectangle-shaped airholes with various widths $W_{1}$ and $W_{2}$ (where $W_{1} + W_{2} = 0.5a$, with $a$ representing the period), separated by half a period (0.5$a$) as shown in the inset of central panel in Fig. 1b. The $W_{1}$ (or $W_{2}$) airholes are arranged centrally in the unit cell to maintain inversion symmetry, resulting in quantized Zak phase of either 0 or $\pi$\cite{ota2018topological,xiao2014surface}. Figure 1b shows the calculated transverse-electric (TE) like band diagrams of PhC unit cell for different configurations: $W_{1} = 0.15a$ (left panel, with $W_{1}$ airhole centered), $W_{1} = 0.25a$ (middle panel), and $W_{2} = 0.35a$ (right panel, with $W_{2}$ airhole centered). For $W_{1} = 0.25$, the two airholes are identical, forming gapless bands with Dirac points at the zone edge ($k = \pi/a$). Meanwhile, the PhCs with $W_{1} = 0.15a$ and $W_{2} = 0.35a$ airhole centered possess same band diagrams, and both the first and second photonic bandgap appear, as shown in the left and right panels of Fig. 1b. In contrast to reported 1D topological PhC nanolaser by using the edge-state mode in the lowest-energy bandgap\cite{ota2018topological}, the topological edge state within the second bandgap is implemented in this work, enabling better tolerance to fabrication errors and improved mechanical stability due to larger lattice constant and size of airholes. The Zak phase of band $A1$ ($A2$) for the PhC with $W_{1} = 0.15a$ airhole centered is $\theta$$_{A1}^{Zak}$ $=$ 0 ($\theta$$_{A2}^{Zak}$ $=$ $\pi$), indicating topologically trivial (nontrivial) character. While the corresponding band $B1$ ($B2$) is topologically nontrivial (trivial) with $\theta$$_{B1}^{Zak}$ $=$ $\pi$ ($\theta$$_{B2}^{Zak}$ $=$ 0) for $W_{2}$ $=$ 0.35$a$ airhole centered. Thus, a topological nanocavity (Fig. 1e) can be created by splicing two PhC nanobeams with distinct Zak phases, producing a robust topologically protected edge mode within the second bandgap at the interface.

The resonant frequencies and Q-factors of the edge mode within second bandgap for different $W_{1}$ values are illustrated in Fig. 1c, with the structural parameters $a =$ 845 nm, $H = $810 nm, $h =$ 526 nm and six periods of PhC on each side. The Q-factor reachs a peak value of around 2000 at $W_{1} =$ 0.23.
To achieve high Q-factor and small $V_{m}$, the topological nanocavity is optimized by tapering the four nearest lattices ($a_{i+1} = a_{i} + d_{a}$, $i = 1 - 3$, and $a = a_{4} + d_{a}$) at the interface as shown in Fig. 1f, which is similar to high-Q tapered PhC defect nanocavities\cite{zhou2021single,deotare2009high}. Figure 1d shows the evolution of Q-factor and $V_{m}$ with various $d_{a}$ while keeping $W_{1} = 0.22a$. The Q-factor increases significantly and peaks at $d_{a}$ = 0.02$a$ with a value of approximately 5.6 $\times$ 10$^{4}$. The $d_{a}$ values also essentially affect the optical penetration depth into the PhC, resulting in an exponential-like reduction in $V_{m}$ from around 1.9 $(\lambda/n)^{3}$ at $d_{a} = 0$ to a wavelength-scale $V_{m}$ of approximately 0.65 $(\lambda/n)^{3}$ at $d_{a} = 0.08a$. The remarkable light localization with reduced $V_{m}$ is validated by the simulated mode profiles, as shown in Fig. 1e and Fig. 1f. The electric field ($E_{y}$) profile of the edge mode without tapered manipulation (Fig. 1e) exhibits a broad field distribution for $W_{1} = 0.22a$, while the tapered structure achieves ultra-compact field localization at the interface (Fig. 1f). Accordingly, we take $d_{a} = 0.05a$ for the optimal design with both a high Q-factor (2.9 $\times$ 10$^{4}$) and a small $V_{m}$ (0.68 $(\lambda/n)^{3}$). Figures 1(g) and 1(h) display the simulated $V_{m}$, Q-factor, and resonant wavelength of edge-state mode for various height ($H$) of the tapered topological nanocavity ($W_{1} = 0.22a$, $h = $583 nm and $d_{a} = 0.05a$). Both an ultra-small $V_{m}$ and a high Q-factor can be simultaneously achieved within the target wavelength range. Besides, the resonant wavelength shows near-linear dependence on $H$, allowing precise wavelength control by adjusting nanobeam height (Fig. 1h). The designed tapered topological PhC nanocavity, with its high Q-factor and ultra-small $V_{m}$, is promising for constructing ultra-compact topological rainbow nanolaser. To realize nanoscale rainbow emission, multiple single-mode tapered topological nanocavities are tightly integrated into a single PhC nanobeam. The middle panel of Fig. 1i shows a top-view SEM image of the fabricated ultra-compact 1D topological rainbow nanolaser, assembled by alternately splicing topological nontrivial (orange) and trivial (blue) PhC regions. The period ($a$) and height ($H$) of each PhC lattice are shown in the bottom panel in Fig. 1i. An effective index modulation by gradually increasing $H$ along the PhC nanobeam from left to right is applied, enabling controllable lasing wavelength and frequency spacing of the rainbow nanolaser. The calculated electric field ($E^{2}$) profiles of the spatially separated trapped topological mode at various wavelengths (1550 nm, 1570 nm, and 1590 nm) are shown in the top panel of Fig. 1(i), confirming the rainbow trapping effect in the designed structure.

\begin{figure}[t]
	\centering
	\includegraphics[width=0.93\linewidth]{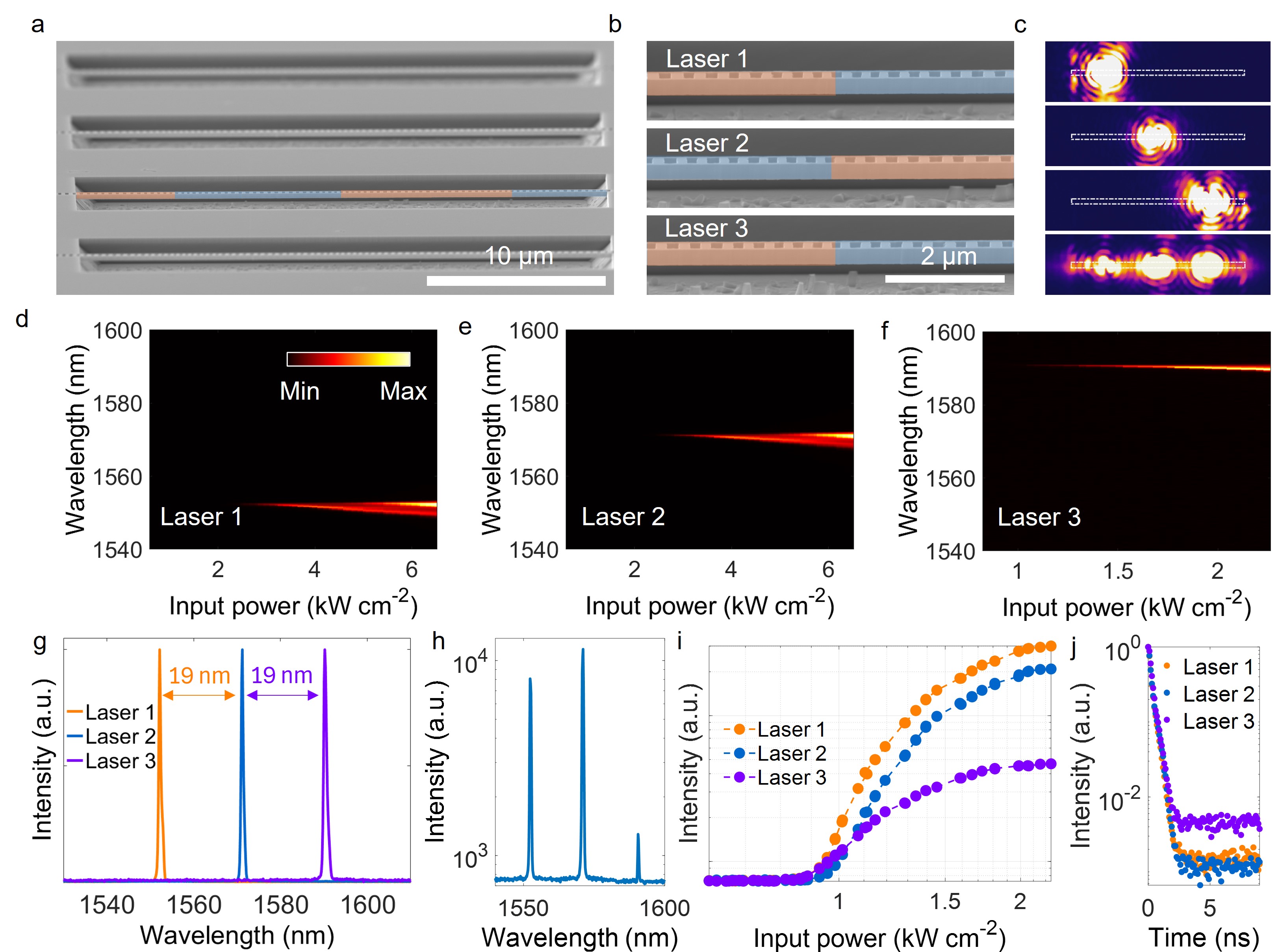}
	\caption{\textbf{Ultra-compact 1D Topological rainbow nanolaser}. (a) The tilted SEM image of fabricated topological rainbow nanolaser array. (b) The SEM images of three integrated topological PhC nanocavities within a rainbow nanolaser. (c) The near-field images of the lasing mode measured above the threshold by individually pumping (top three panels) and uniform pumping (bottom panel) method. The white dashed line outlines the fabricated device. (d)-(f) Measured lasing spectra at various input powers by selectively pumping three nanocavities of the topological rainbow nanolaser. Lasing spectra of the 1D topological rainbow nanolaser collected by selective (g) and uniform (h) excitation method, respectively. (i) The measured $\textit{L}-\textit{L}$ curves of the three lasing peaks. The thresholds are in the same order of $\sim$ 0.9 kW cm$^{-2}$. (j) TRPL spectra of the three lasing peaks measured above thresholds, with fitted lifetimes ($\tau$$_{lasing}$) of 0.28 ns, 0.28 ns and 0.33 ns for Laser 1, Laser 2, and Laser 3, respectively.}
	\label{fig3}
\end{figure}

To verify the optimized topological nanocavity, we initially fabricate single-mode tapered topological nanolasers by using InGaAsP multi-quantum wells (MQWs) as gain materials. Figures 2a and 2b display SEM images of the fabricated devices and the central tapered cavity region, respectively.
The measured spectra of a tapered topological PhC nanolaser with $a =$ 815 nm, $W_{1} =$ 0.22 , $h =$ 495 nm, and $H$ = 810 nm are shown in Fig. 2c, revealing robust single-mode emission. The normalized above-threshold spectra (solid curve) and spontaneous emission spectra (dash curve) measured from an unpatterned region are shown in the inset, illustrating significant suppression of the spontaneous emission after threshold.
The corresponding light-in/light-out ($\textit{L}-\textit{L}$) curve and the linewidth are shown in Fig. 2d, providing clear evidences of lasing operation through a distinct kink in the $\textit{L}-\textit{L}$ curve and linewidth narrowing effect. The estimated lasing threshold of approximately 1.6 kW cm$^{-2}$ (average power 400 nW) is about two orders of magnitude lower than reported C-band topological nanolasers\cite{kim2020multipolar}. 
A Gaussian curve fit to the spectra measured just below threshold (Fig. 2e) indicates a linewidth of approximately 0.4 nm and an experimental Q-factor 
$\sim$ 3900. The topological edge-state mode is confirmed by the near-field optical profiles, as shown in Fig. 2f.
Strong speckle patterns appear above the threshold, attributed to the high coherence of the emission. Figure 2g shows collected spectra under various input powers, with a notable $\sim$ 0.7 nm blueshift of the lasing peak as input power increases, attributed to the band-filling\cite{mylnikov2020lasing,nguyen2017hybrid} and carrier plasma effect\cite{ide2004lasing}. To further understand the lasing operation, time-resolved photo-luminescence (TRPL) measurements are performed at room temperature to investigate carrier dynamics. Figure 2h displays normalized TRPL spectra for the unpatterned region (blue), the topological nanolaser pumped slightly below (yellow) and above (green) threshold, corresponding to spontaneous, amplified spontaneous (ASE) and stimulated emission, respectively. 
A clear transition from spontaneous emission with a long lifetime ($\tau_{spe} = 4.01$ ns) to ASE ($\tau_{ASE} = 0.69$ ns), then to coherent stimulated emission with a shorter lifetime ($\tau$$_{lasing} =$ 0.26 ns) is observed, presenting an additional key signature of lasing. In addition, the normalized lasing spectra and corresponding thresholds of representative tapered topological nanolasers with various $a$ and $h$ are respectively shown in Fig. 2i and Fig. 2j, revealing a robust widely tunable wavelength range of approximately 80 nm. Variations in cavity structure and fabrication lead to different thresholds for topological nanolasers with similar wavelengths, as indicated in Fig. 2j.

\begin{figure}[t]
	\centering
	\includegraphics[width=0.93\linewidth]{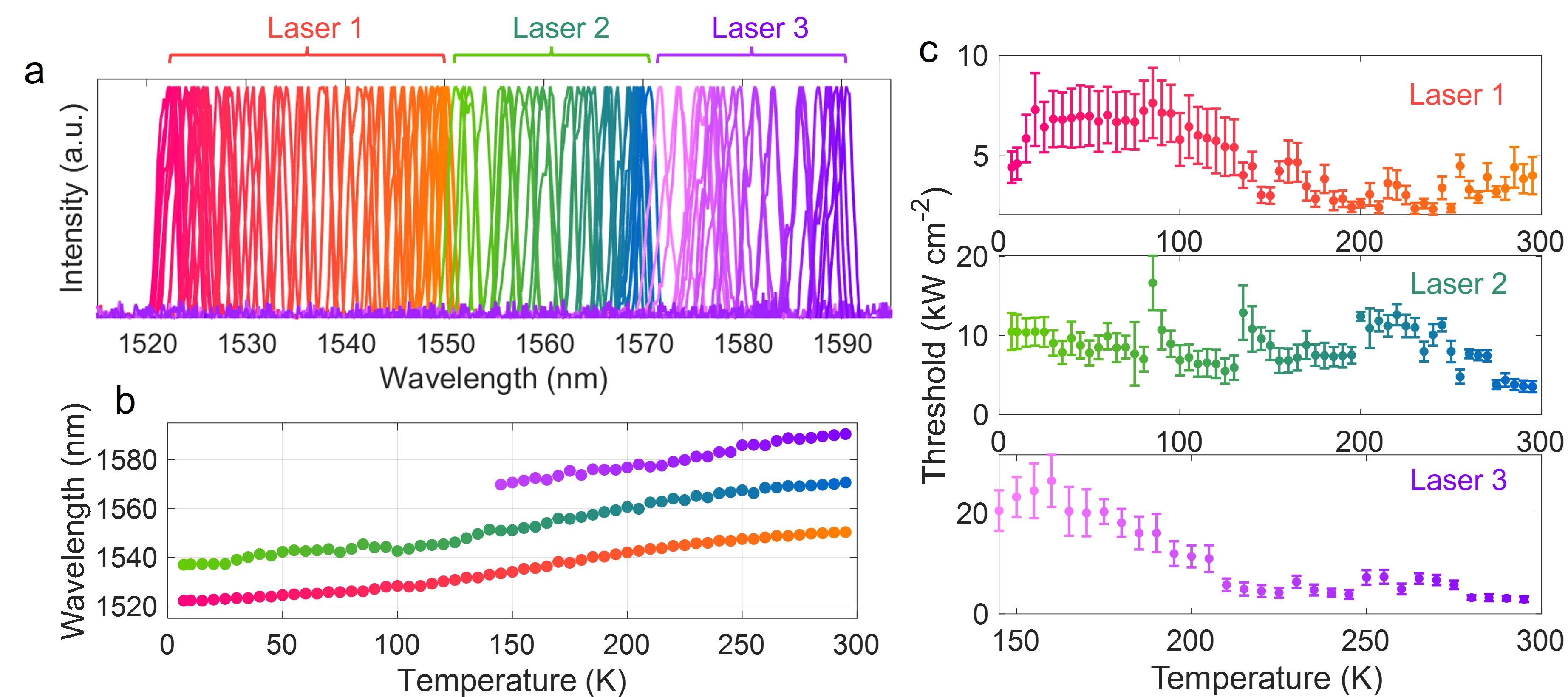}
	\caption{\textbf{Temperature-dependent behavior of the 1D topological rainbow nanolaser}. The measured normalized lasing spectra (a), lasing wavelengths (b), and thresholds (c) of the 1D topological rainbow nanolaser at various temperatures. The data are shown in three color groups, representing experimental results for Laser 1, Laser 2, and Laser 3, respectively.}
	\label{fig3}
\end{figure}

Afterwards, the ultra-compact 1D topological rainbow nanolaser is fabricated based on the designed tapered topological nanocavity. Figures 3a and 3b show false-color SEM images of fabricated 1D topological rainbow nanolasers and an enlarged view of three  integrated nanocavities with distinct lasing wavelengths (Laser 1, Laser 2, and Laser 3). Benefiting from the wavelength-scale $V_{m}$, these closely integrated topological nanolasers are expected to operate independently, which is confirmed by the near-field profiles. As shown in the top three panels of Fig .3c, the measured optical profiles of the three topological edge modes with discrete frequencies are spatially separated and highly localized at the interfaces. The spectra of the 1D topological rainbow nanolaser is characterized by selectively pumping three nanocavities. Figures 3d$-$3f present the corresponding measured spectra under various input powers for the three lasing peaks (Laser 1, Laser 2, and Laser 3) of a topological rainbow nanolaser with $a = 845$ nm and gradually increased $H$ (Fig. 1i). The normalized lasing spectra are shown in Fig. 3g. An uniform interval of approximately 19 nm between the three peaks (1552 nm, 1571 nm and 1590 nm) is observed, which closely matches the simulation results (Fig. 1i). Furthermore, the 1D topological rainbow spectra can be excited by using a strip-shaped pumping beam across the entire device. The corresponding measured field profile of the 1D topological rainbow nanolaser is shown in the bottom panel of Fig .3c and the lasing spectra is shown in Fig. 3h, indicating three nanolasers are simultaneously excited.
The first two nanolasers exhibit output intensities roughly ten times that of the third nanolaser (Fig. 3h), verified from the collected $\textit{L}-\textit{L}$ curves with a higher slope efficiency as shown in Fig. 3i. The discrepancy also exists in the carrier lifetime. The carrier lifetime above threshold of the third nanolaser ($\tau_{lasing} = 0.33$ ns) is slightly longer than that of the first two nanolasers ($\tau_{lasing} =$ 0.28 ns) (Fig. 3j). Such differences may result from the lower Q-factor and gain at longer wavelengths.

\begin{figure}[t]
	\centering
	\includegraphics[width=0.93\linewidth]{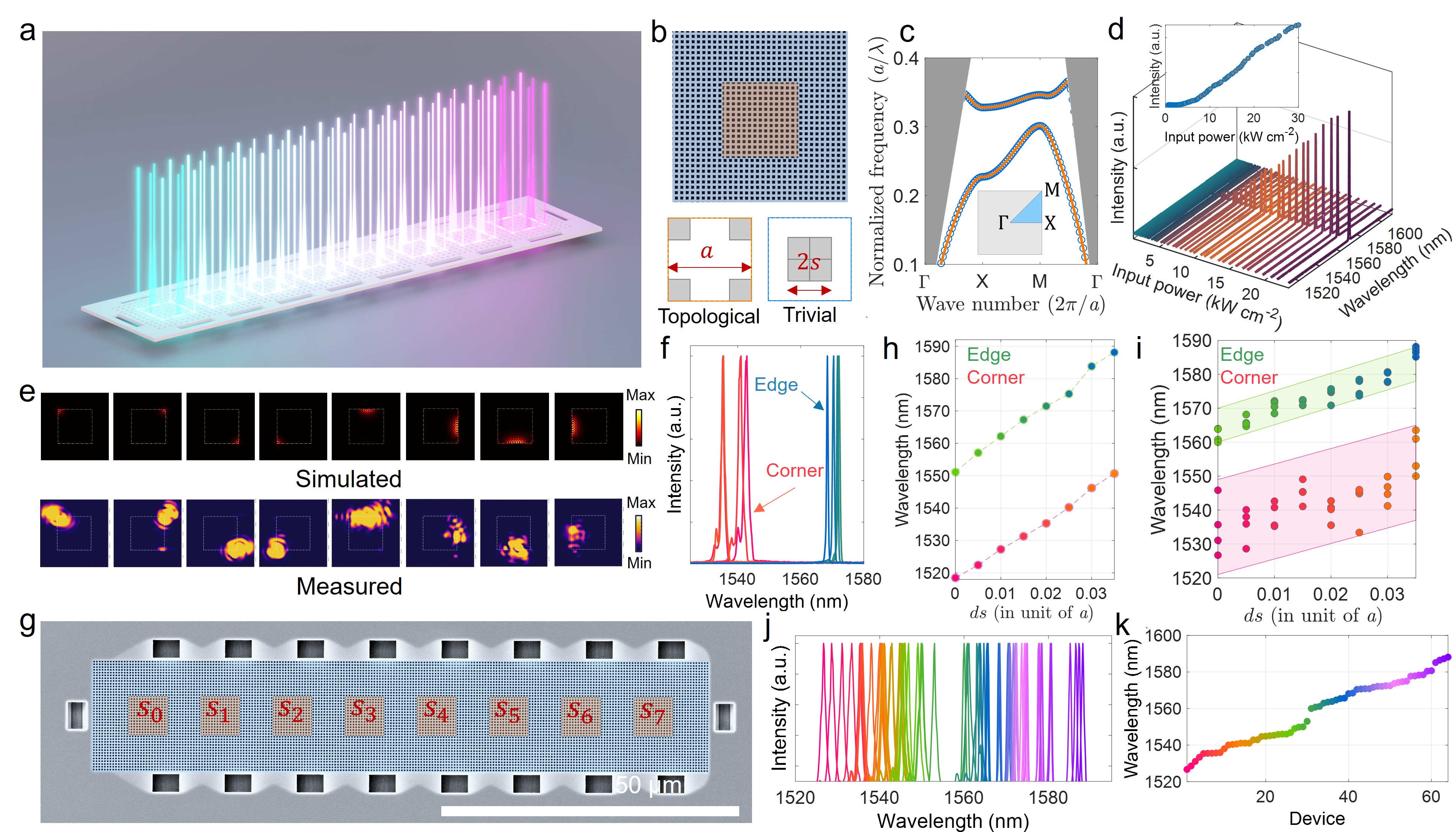}
	\caption{\textbf{Ultra-compact 2D topological rainbow nanolaser}. (a) Schematic of the 2D topological rainbow nanolaser with 64 distinct lasing wavelengths. (b) SEM image of the fabricated topological nanocavity. The unit cells of the topologically nontrivial and trivial PhC are shown in the bottom panel. (d) Calculated TE-like band diagram of the topologically nontrivial (yellow line) and trivial PhC (blue circle) with $a =$ 470 nm and $s = 0.29a$. (d) Measured spectra of a topological edge-state nanolaser with $a =$ 470 nm and $s = 0.28a$. The threshold is estimated to be around 5 kW cm$^{-2}$ from the $\textit{L}-\textit{L}$ curve shown in the inset. (e) Simulated (top panel) and measured (bottom panel) lasing optical profiles for four topological corner modes and four topological edge mode of the 2D topological nanocavity. The white dashed line illustrates the boundary of topologically distinct domains. (f) Measured edge-state and corner-state lasing spectra of a topological nanocavity with $a =$ 470 nm and $s = 0.28a$. (g) Top-view SEM image of a fabricated 2D topological rainbow nanolaser. Simulated resonant wavelengths (h) and experimental lasing peaks (i) for topological edge and corner modes by varying side lengths ($s$) of the topologically nontrivial PhC with $a = 470$ nm, $s_{0} = 0.29a$, and $ds = s_{0} - s_{i}$ (for $i =$ 0 to 7). Rainbow spectra (j) and the associated lasing wavelengths (k) of the fabricated 2D topological rainbow nanolaser.  }
	\label{fig3}
\end{figure}

Temperature-dependent behavior of the fabricated 1D topological rainbow nanolaser is characterized from 7 K to room temperature. Figure 4a shows the normalized lasing spectra of the device at various temperatures, with three distinct color groups representing the three lasing peaks. A continuous blue-shift in lasing wavelengths is observed as the temperature decreases, attributed to the temperature-induced changes in the refractive index and gain spectra. Meanwhile, the temperature-dependent emission exhibits robust single-mode operation with precisely tunable wavelengths, fully covering the 19 nm intervals between the three lasing peaks. Therefore, a continuously broad rainbow-like spectra spanning of approximately 70 nm is achieved within the 1D topological rainbow nanolaser. The detailed wavelengths for the 1D topological rainbow nanolaser are shown in Fig. 4b. The first two peaks (Laser 1 and Laser 2) exhibit robust lasing from 7 K to 300 K with wavelength ranging from 1522$-$1550 nm and 1537$-$1571 nm, respectively. The third peak (Laser 3) is tuned from 1570 nm to 1590 nm as the temperature increases from 145 K to 300 K. Lasing is not observed for the third nanolaser below 145 K, mainly due to the significant mismatch and limited overlap between the resonant wavelength and gain spectra at lower temperatures. The corresponding thresholds of the 1D topological rainbow nanolaser are shown in Fig. 4c. Generally, the nonradiative recombination rates are dramatically suppressed and the threshold is expected to decrease rapidly at low temperature, especially for PhC nanolasers with high surface-to-volume ratios. However, the experimental thresholds of the topological PhC rainbow nanolaser show a slight increase as temperature decreases, deviating from the usual trend. This discrepency may predominantly attributed to an increased mismatch between the lasing wavelength and gain spectra, which requires higher input power to reach stimulated emission.

Additionally, by implementing the same strategy, we further demonstrate an ultra-compact 2D topological rainbow nanolaser with 64 separated lasing peaks through integrating spatially distributed 1D edge modes and 0D corner modes. Figure 5a displays a schematic of the designed 2D topological rainbow nanolaser. The top panel of Fig. 5b shows a zoomed-in SEM image of one topological nanocavity. A topologically nontrivial PhC region (orange) is embedded within a trivial PhC domain (blue), trapping four robust topologically protected 1D edge-state modes and four 0D corner-state modes with ultra-small $V_{m}$ at the interfaces \cite{kim2020multipolar}. The unit cell of the considered topologically distinct PhC, depicted in the lower panel of Fig. 5b, consists of four square-shaped airholes with side lengths $s$ and lattice constant $a$. The topologically nontrivial and trivial PhC structures share an identical opened gap band diagram while presenting various topologies\cite{zhou2022monolithically}. Figure 5d illustrates measured lasing spectra for one of the integrated topological edge-state nanolasers, presenting robust single-mode operation in a wide range of input powers. For the designed topological cavity, the topological edge-state modes and corner-state modes can be selectively excited by scanning the pumping spot along the edges or corners of the interfaces, as evidenced by the bright coherent emission at these locations (Fig. 5e). The four corner-state nanolasers exhibit similar wavelengths around 1540 nm, whereas the four edge-state nanolasers have longer wavelengths near 1570 nm for a topological nanocavity with $a =$ 470 nm and $s =$ 0.28$a$ (Fig. 5f). Benefiting from the minor fabrication imperfections, both the edge modes and corner modes exhibit a degree of wavelength fluctuation, offering a potential scheme to realize multi-wavelength rainbow spectra in a limited area. To achieve rainbow-like emission, eight topological nanocavites with gradually modulated structural parameters are densely integrated in a single device, confining 64 spatially separated topological modes with finely tuned lasing wavelengths. Figure 5g presents a SEM image of the fabricated 2D topological rainbow nanolaser with an ultra-small footprint of approximately 0.002 mm$^{2}$. Eight topological nontrivial domains with various structural paramters ($s_{0} = 0.29a$, $s_{i} - s_{i+1} = 0.05a$ for $i =$ 0 to 7) are embedded within a trivial domain of uniform side length $s_{0}$. By modulating the side lengths of the topological domain, the corresponding simulated resonant wavelengths of the edge-state and corner-state modes are linearly tuned as indicated in Fig. 5h. In this configuration, the wavelength gap ($\sim$ 30 nm) between edge mode and corner mode can be filled and cover the entire gain region by splicing edge and corner nanolasers. For the fabricated rainbow nanolaser, we observe robust single-mode emission from all of the integrated 32 topological edge nanolasers and 32 topological corner nanolasers. The measured lasing peaks of the integrated topological modes agree well with simulated data (Fig. 5i), exhibiting a trend of red-shift with smaller $s$ due to an increased effective refractive index. Figures 5j and 5k illustrate the measured normalized rainbow-like spectra and lasing peaks of the fabricated 2D topological rainbow nanolaser, spanning almost the entire C-band with partial extension into the L-band.

\section*{Conclusion}

In summary, we demonstrate the design and experimental realization of nanoscale rainbow light trapping for lasing emission in the 1550 nm telecom band. Ultra-compact 1D and 2D topological rainbow nanolasers are presented with diffraction-limited $V_{m}$ and ultra-low lasing threshold. The 1D topological rainbow nanolaser is constructed by splicing three tapered topological nanocavities, exhibiting rainbow-like emission with an uniform wavelength interval of approximately 19 nm modulated by the height of PhC nanobeam. Meanwhile, a continuously wavelength-tunable topological rainbow nanolaser with a broad spectral range of approximately 70 nm (1520$-$1590 nm) is achieved through controlling the temperature. Additionally, by means of splicing 2D topological nontrivial and trivial domains, we densely integrate 64 topologically protected resonant modes within an extremely compact area of around 0.002 mm$^{2}$, producing a rainbow spectrum with a wavelength span of 70 nm. We believe that the demonstrated robust topological rainbow nanolaser will pave the way for developing ultra-compact, fast, and efficient light sources for next-generation densely integrated nanophotonic circuits, particularly for on-chip wavelength-division multiplexing and intra-chip optical interconnects.


\section*{Methods}

\subsection*{Simulation} 
The photonic band diagrams were calculated in a 3D model through the plane-wave expansion method\cite{johnson2001block}. Commercial software (Lumerical) was performed to simulate the field profile, Q-factor and $V_{m}$. The $V_{m}$ of the topological nanocavity was calculated by $V_{mode} =\frac{\int\varepsilon(\textbf{r}) E(\textbf{r})^{2}dV}{max(\varepsilon(\textbf{r}) E(\textbf{r})^{2})}$, where $\varepsilon(\textbf{r})$ and $E(\textbf{r})$ represent the local permittivity and electric field intensity, respectively. 

\subsection*{Device fabrication} 
The topological rainbow nanolasers were fabricated using a 260-nm-thick InGaAsP/1.3-$\mu$m-thick InP/50-nm-thick InGaAs/InP substrate wafer. The 260-nm-thick active layer contains six compressively strained InGaAsP MQWs ($\lambda$ = 1.55 $\mu$m) that serve as the gain medium. The 1.3-$\mu$m-thick InP and 50-nm-thick InGaAs layer act as sacrificial and etch stop layer, respectively. The topological rainbow PhC structure was defined by electron beam lithography on a 300-nm-thick PMMA resist film coated on the wafer. The pattern was subsequently transferred through the 90-nm-thick SiO$_{2}$ hard mask and into the gain layer via plasma dry etching. Afterwards, the remaining PMMA layer was removed by O$_{2}$ plasma, and the residual SiO$_{2}$ hard mask was eliminated in a buffered oxide etching solution. Finally, the suspended topological rainbow nanolasers were formed by selectively undercutting the InP sacrificial layer using diluted HCl solution. 

\subsection*{Optical measurement}

The fabricated devices were optically pumped at room temperature in a custom-built micro-photoluminescence system. A 10 ns 632 nm pulsed laser operating at an 800 kHz repetition rate was used as the excitation source to avoid thermal effects. The pumping laser spot, with a diameter of approximately 2 $\mu$m, was focused using a high-NA (0.9) 100$\times$ objective lens (MPlan N, Olympus) and precisely positioned at the center of the nanocavity region with piezo-electric nanopositioners (New Focus). The average input powers were directly measured by using a high resolution photodetector (S120C, Thorlabs), and the peak input powers are calculated by considering the duty cycle and spot size of input laser beam. The emission spectra were collected from the top through the same objective lens and analyzed by the spectrometer (SpectraPro HRS-500, Princeton Instruments) with a liquid nitrogen-cooled infrared InGaAs detector (PyLoN-IR, Princeton Instruments). The near-field images were measured using a thermoelectrically cooled InGaAs camera (KnightSWIR-130). Long-pass filters were used to prevent the excitation laser from reaching the detectors. To simultaneously excite topological rainbow nanolasers, two cylindrical lenses were used to create a strip-shaped pump beam that covered the entire device. The carrier lifetime measurements were performed using a TCSPC system, consisting of a time controller (ID1000, ID Quantique), a single photon detector (IDQube-NIR, ID Quantique) and a 407 nm picosecond pulsed diode laser (PiL040-FS, NKT Photonics). The carrier lifetimes are extracted from fitting the experimental TRPL data using a standard bi-exponential component function ($\textit{I}(\textit{t}) = $\textit{A}$_{1}$exp(-\textit{t}/$\tau$$_{1}$)$ +$ \textit{A}$_{2}$exp(-\textit{t}/$\tau$$_{2}$)), where $\tau$$_{1}$ and $\tau$$_{2}$ are the fast and slow decay components, respectively. For temperature-dependent measurements, the devices were mounted in a liquid helium cryostat (Montana Instruments) with the temperature controlled from 7 K to 300 K.

\section*{Acknowledgements} 
This work is supported by the National Natural Science Foundation of China under Grants No.62304080, 62074018 and 62174015, the Guangdong Basic and Applied Basic Research Foundation under Grant No. 2024A1515010802, the Science and Technology Projects in Guangzhou under Grant No. 2024A04J3683, and the Fundamental Research Funds for the Central Universities under Grant No. 2023ZYGXZR068. T.Z. acknowledges the startup funds from South China University of Technology. The authors would like to thank Zhi Xu from Songshan Lake Materials Laboratory for the electron beam lithography processing help.

\section*{Author contributions}

T.Z. conceived the idea and developed the theoretical concept. F.T. and W.H. fabricated the devices. Y.W. carried out the optical measurements. S.G. and T.Z. performed numerical calculations. T.Z. wrote the paper with input from all coauthors. X.F. and T.Z. supervised the project.

\section*{Competing interests} 

The authors declare no competing interests.

\section*{Data availability} 

The data are available from the authors upon reasonable request.

\bibliography{sample}

\end{document}